\documentclass[12pt,a4paper]{article}
\usepackage{graphics}
\usepackage{cite,a4wide}
\newcommand{\beq}{\begin{equation}}
\newcommand{\eeq}{\end{equation}}
\newcommand{\bea}{\begin{eqnarray}}
\newcommand{\eea}{\end{eqnarray}}

\renewcommand{\b}{\beta}
\renewcommand{\a}{\alpha}
\renewcommand{\ni}{\noindent}

\newcommand{\m}{\mu}

\newcommand{\s}{\sigma}

\newcommand{\oh}{\frac{1}{2}}

\newcommand{\dg}{\dagger}
\newcommand{\non}{\nonumber}

\newcommand{\ra}{\rightarrow}

\begin{document}

\hfill July 1998

\begin{center}

\vspace{32pt}

  { \Large \bf Evidence for a Center Vortex Origin \\ 
                  of the Adjoint String Tension} 

\end{center}

\vspace{18pt}

\begin{center}
{\sl M. Faber${}^a$, J. Greensite${}^b$,
and {\v S}. Olejn\'{\i}k${}^c$}

\end{center}

\vspace{18pt}

\begin{tabbing}

{}~~~~~~~~~~~~~~~~~~~~\= blah  \kill
\> ${}^a$ Inst. f\"ur Kernphysik, Technische Universit\"at Wien, \\
\> ~~A-1040 Vienna, Austria.  E-mail: {\tt faber@kph.tuwien.ac.at} \\
\\
\> ${}^b$ The Niels Bohr Institute, DK-2100 Copenhagen \O, \\
\> ~~Denmark.  E-mail: {\tt greensite@nbivms.nbi.dk} \\
\\
\> ${}^c$ Institute of Physics, Slovak Academy of Sciences, \\
\> ~~SK-842 28 Bratislava, Slovakia.  E-mail: {\tt fyziolej@savba.sk}

\end{tabbing}

\vspace{18pt}

\begin{center}

{\bf Abstract}

\end{center}

\bigskip

    Wilson loops in the adjoint representation are evaluated on cooled
lattices in SU(2) lattice gauge theory.  It is found that the string 
tension of an adjoint Wilson loop vanishes, if the loop is evaluated in 
a sub-ensemble of configurations in which no center vortex links the loop.  
This result supports our recent proposal that the 
adjoint string tension, in the Casimir-scaling regime, can be
attributed to a center vortex mechanism. 

\vfill

\newpage

   The center vortex theory of quark confinement has received 
increasing support, in the last year or so, from numerical simulations
conducted by our own group \cite{PRD98,Zako}, by Langfeld et al.\
\cite{LR}, and by Kov\'{a}cs and Tomboulis \cite{TK} 
(a brief summary of our own results can be found at the end of 
ref.\ \cite{PRD98}).  Despite these successes, the existence and approximate 
Casimir scaling of the adjoint-representation string tension would seem
to be problematic for the vortex theory, since adjoint Wilson loops are
insensitive to the gauge-group center \cite{Cas1,lat96}.  This issue was 
addressed some months ago in ref. \cite{Us}, where we argued that the 
adjoint (and higher) representation string tensions can \emph{also} be 
attributed to a center vortex mechanism.  In this letter we present 
some numerical data in support of that argument.

   Let us begin by recalling that, while the quark-antiquark string 
tension depends on the group representation of the heavy-quark color charges, 
this dependence is different in two separate regimes.  In the 
{\bf Casimir-Scaling Regime}, extending from the onset of confinement to
the onset of color screening, the string tension $\s_r$ for the
$r$-representation seems (from numerical experiments \cite{Cas}) to be
very roughly proportional to the eigenvalue of the quadratic 
Casimir; e.g. for the SU(2) gauge group
\beq
       {\s_j \over  \s_{1/2}} \approx {4\over 3} j(j+1)
\eeq
In particular, the SU(2) adjoint string tension is $\approx 8/3$ the 
fundamental string tension at intermediate distances.  
However, for sufficiently large quark separation,
the quark color charges must be screened by gluons to the lowest 
representation with the same transformation properties under the $Z_N$
subgroup.  This is the {\bf Asymptotic Regime} where, in the SU(2) case, 
\beq
       \s_j = \left\{ \begin{array}{cl}
                       \s_{1/2} & j=\mbox{half-integer} \cr
                           0    & j=\mbox{integer} \end{array} \right.
\eeq
In the $N\ra \infty$ limit, color-screening is suppressed, and Casimir-scaling
is exact out to infinite quark separations.  But even at $N=2$ there seems to 
be a Casimir regime of some finite extent.

    Explaining the behavior of the string tension in both the
asymptotic \emph{and} Casimir scaling regimes is a 
major challenge for any theory of confinement.  In particular, in the
case of the vortex theory, creation of a center vortex linking loop $C$
multiplies the value $W_F(C)$ of the fundamental representation 
Wilson loop, in an SU(N) gauge theory, by an element of the center,
i.e. $W_F(C) \ra z W_F(C)$, where $z=\exp[2\pi i n/N] \in Z_N$.  The 
vortex theory attributes the area-law
falloff of $<W_F(C)>$ to fluctuations in the number of vortices linking
the loop.  However, since $W_A(C) = \Bigl| W_F(C) \Bigr|^2 -1$,
adjoint loops would seem to be unaffected by the presence of center
vortices.  This insensitivity to vortices leads to the correct asymptotic 
result that there is no adjoint string tension at large distance scales, 
but then how does one explain the existence of an adjoint tension in the 
Casimir regime?

    Our proposed answer to this question in ref.\ \cite{Us} (see also
related considerations by Cornwall \cite{Corn}) begins by 
noting that center vortices are surface-like objects, with a finite 
thickness (a ``core'') on the order of magnitude of the 
confinement scale.  Only outside this
finite core can the effects of the vortex be simply represented by a 
discontinuous gauge transformation.  The statement that adjoint loops are 
unaffected by center vortices must, therefore, be qualified: it is only true 
providing the vortex core nowhere overlaps the loop perimeter.  This leads
us to ask:  What is the effect of vortices on adjoint loops whose
minimal area is comparable to, or smaller than, the cross-sectional
area of the vortex core?  For such loops, the overlap of the vortex core 
with the loop perimeter cannot be disregarded.

\begin{figure}[h]
\centerline{\scalebox{.5}{\rotatebox{270}{\includegraphics{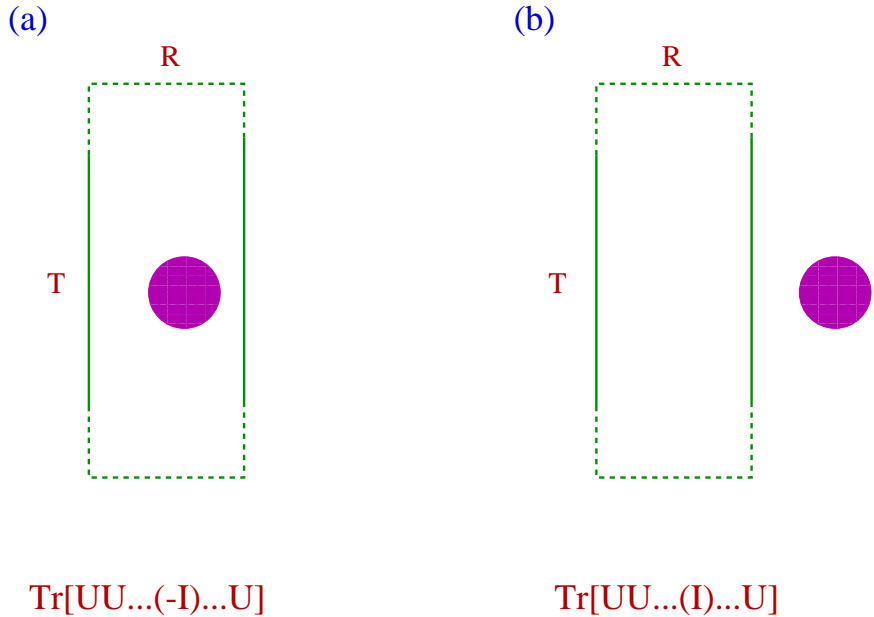}}}}
\caption{Vortex core (shaded region) intersecting the plane of an $R\times T$
Wilson loops.  With no core/perimeter overlap, the effect of the vortex
is the insertion of a center element in the link product.}
\label{vortin}
\end{figure}

\begin{figure}[h]
\centerline{\scalebox{.5}{\rotatebox{270}{\includegraphics{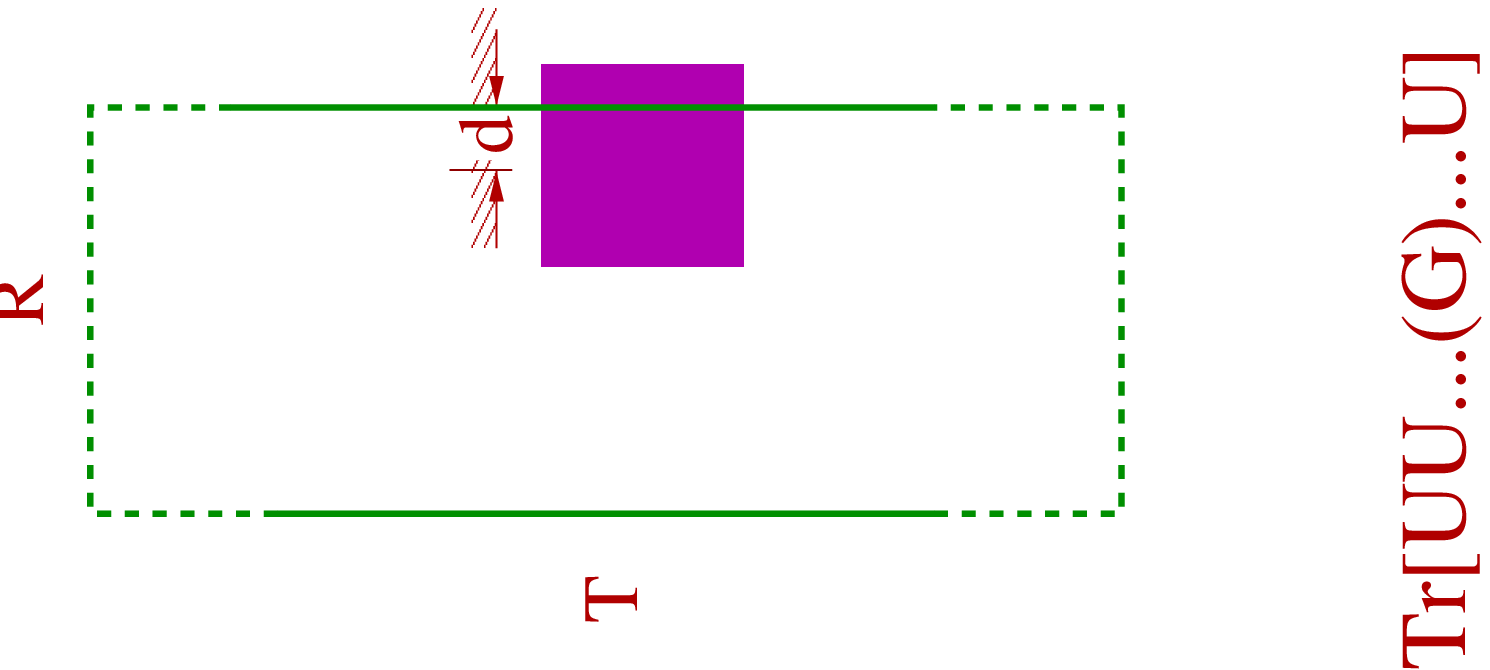}}}}
\caption{Overlap of the vortex core with the loop perimeter.  The effect
of the vortex is represented by insertion of a group element $G$ in the
link product.}
\label{vortolap}
\end{figure}

    In ref.\ \cite{Us} we studied the effects of the overlap in the
context of a simple model. Consider the case of an $R\times T$ loop in
SU(2) lattice gauge theory, $T \gg R$, lying in the x-t plane. 
If the vortex core pierces the minimal area of a loop, but does not overlap the
loop perimeter (Fig. 1a), then the effect of the vortex is simply the insertion
of a center element ($-I$ in this case) somewhere in the product of the
link variables.  If we then imagine displacing
the vortex so that the cross-section of the core, in the plane of the loop,
lies entirely outside the loop (Fig. 1b), then the center element $-I$ is 
replaced by $+I$.  Finally, if the vortex core overlaps the loop perimeter, as 
shown in Fig.\ \ref{vortolap}, we will assume that its effect on the loop can 
be represented by insertion of a group element $G$ into the product of
link variables

\beq
      G = S \exp\Bigl[i\a_R(x) {\s_3\over 2} \Bigr] S^\dg 
                ~~~~~~~~~~~~~~~0\le \a_R(x) \le 2\pi
\eeq
which interpolates smoothly between the limiting cases $G=\pm I$ at
$\a_R = 0$ and $\a_R=2\pi$ respectively.  Here $S$ is some (randomly 
distributed) SU(2) group element, and $\a_R(x)$ depends on the fraction of 
the core cross-section contained in the minimal area of the loop. Position
$x$ locates the middle of the vortex where it pierces the plane of the
loop.  The trace of the link product can be taken in any representation.  
We also define  
\beq
  f \equiv \mbox{probability for the middle of a vortex to 
                     pierce a plaquette}
\eeq
If both the positions $x$ and group orientations $S$ of 
vortices piercing the plane of the loop are uncorrelated, one finds for 
the vortex-induced potential between heavy color charges in 
representation $j$ \cite{Us}
\bea
      V_j(R) &=& 
          - \sum_{n=-\infty}^{\infty}\ln\{(1-f)
          + f{\cal{G}}_j[\a_R(x_n)] \}
\non \\
      {\cal{G}}_j[\a] &=& 
         {1\over 2j+1} \sum_{m=-j}^{j} \cos(\a m)
\eea
where $x_n=(n+\oh)a$, with $a$ the lattice spacing, now denotes
a (dual) lattice coordinate on the x-axis. For large quark separations, 
this expression leads to the correct asymptotic form for the string tension
\beq
     \s_j = \left\{ \begin{array}{cl}
               -\ln(1-2f) & j=\mbox{half-integer} \cr
                     0    & j=\mbox{integer} \end{array} \right.
\eeq
while for small $R$, where all $\a_R(x) \ll 2\pi$, we find
\beq
      V_j(R) = \left\{ {f\over 6}\sum_{n=-\infty}^{\infty} 
                 \a_R^2(x_n) \right\} j(j+1)
\label{casprop}
\eeq
which is proportional to the SU(2) quadratic Casimir.  This derivation
of Casimir proportionality at small $R$ generalizes readily from SU(2) 
to SU(N).  

  To actually compute $V_j(R)$ at small and intermediate $R$, even in
this simplified model, one needs to know the function $\a_R(x)$.
Fortunately, this function is constrained to satisfy certain limits, which
gives us a good idea of its shape.  Take the two static charges to lie at
$x=0$ and $x=R$, and
let $d(x)$ be the distance from $x$ to the nearest 
static charge, taken with positive sign if the middle of the vortex is 
outside the $R \times T$ loop (i.e.\ $x<0$ or $x>R$) and 
negative otherwise ($x\in [0,R]$).
The angular variable $\alpha_R(x)$ can only depend on $d(x)$, and
must satisfy the following limits:
\begin{description}
\item{1.} $\a_R(x) \ra 0$ as $d(x) \ra \infty$ 
(vortex far outside the loop)
\item{2.} $\a_R(x) \ra 2\pi$ as $d(x) \ra -\infty$ 
(vortex deep inside a large loop)
\item{3.} $\a_R(x) \ra 0$ as $R \ra 0$ (small loop) 
\end{description}
These conditions are satisfied by, e.g., the simple ansatz
\beq
     \a_R(x) = \pi \left[ 1 - \tanh\left(A d(x) + {B\over R}\right) \right]
\eeq
It then turns out that there does exist a Casimir scaling
region, where $V_j(R)$ is both linear, and approximately proportional
to the quadratic Casimir.\footnote{We find, however, that the deviation from
exact SU(2) Casimir scaling at intermediate distances tends to increase 
with $j$.} The extent of
this Casimir region depends on the choice of parameters $A,~B$.  As 
$R$ increases beyond $1/A$, 
this Casimir scaling behavior goes smoothly over to the behavior 
characteristic of the asymptotic regime.  The details can be found in 
ref.\ \cite{Us}.  An approach having some similarities to ours
(but also differing in important respects, as noted in
\cite{PRD98}), was put forward by Cornwall in ref. \cite{Corn}. 

   While it is very likely that the model described above is an 
oversimplification of the effects of finite vortex 
thickness, it is nonetheless useful in showing
how the adjoint string tension can, in principle, emerge from a center vortex
mechanism.  Moreover, quite apart from the specific details of the model, 
there is one very unambiguous prediction:  If an adjoint loop, located
at some definite position on the lattice, is evaluated in a subensemble of 
gauge field configurations in which no vortex core overlaps the 
loop perimeter, then the string tension of the loop should vanish.  

   Numerically, since the precise boundary of the vortex core is not
sharply defined, it is simpler to study the behavior of the adjoint 
tension under a somewhat weaker restriction, namely, that the middle of any 
vortex core, where it crosses the plane of a loop to be evaluated, is 
exterior to the minimal area of the loop.  This choice of cut in the data 
has been used previously \cite{PRD98,Zako} to define ``no-vortex'' 
loops in the fundamental representation (see below).  As we will see, 
even this weaker restriction seems sufficient to eliminate the adjoint 
string tension.  

   Center vortices are located by a mapping of SU(2) lattice gauge fields onto 
$Z_2$ gauge fields, as explained in refs.\ \cite{PRD98,Zako}.  One begins
by fixing to the ``maximal center gauge,'' defined as the gauge which
maximizes
\beq
        R = \sum_{x,\m} \mbox{Tr}[U_\m(x)]^2
\eeq 
leaving a residual $Z_2$ symmetry.  This gauge brings the link variables
as close as possible, on average, to center elements $\pm I$.  ``Center
projection'' is the mapping
\beq
        U_\m(x) \ra Z_\m(x) = \mbox{signTr}[U_\m(x)]
\eeq  
where the $Z_\m(x)$ transform like $Z_2$ gauge fields under the residual
gauge symmetry.  It is found that the ``thin'' (1-plaquette thick) center 
vortices of the projected configurations $-$ the ``P-vortices'' $-$ locate a
surface in the middle of the ``thick'' center vortices of the full, 
unprojected configurations \cite{PRD98,Zako}.  We can then numerically
evaluate Wilson loops in the unprojected configurations, subject to the 
constraint that a given loop on the lattice is evaluated if and only if a 
definite number $n$ of plaquettes, lying in the minimal area of the loop, are 
pierced by P-vortices. This procedure defines the 
``vortex-limited Wilson loops'' $W^{(n)}(C)$, and in this way we can
study the effects of vortices on unprojected Wilson loops in various 
representations.
 
   For the fundamental representation, it was found in refs.\ 
\cite{PRD98,Zako} that the zero-vortex Creutz ratios $\chi^0_F(R,R)$, 
extracted from zero-vortex Wilson loops $W^0_F(C)$, drop to
zero at sufficiently large $R$.  This is one of the pieces of evidence
in favor of the center vortex theory.  Our proposed vortex mechanism for 
higher representations leads us to predict similar results for 
adjoint-representation, zero-vortex Creutz ratios in the Casimir regime,
where the full adjoint quark potential is roughly linear.  

   Testing this prediction is very straightforward in principle.  The 
practical problem is that the VEVs of higher-representation 
Wilson loops are far smaller than VEVs of the corresponding loops in the 
fundamental representation, and this means that some reduction in noise
due to short-range fluctuations is essential.  A number of
noise-reduction methods exist; we have chosen to use the constrained cooling 
procedure of ref.\ \cite{cool}.  Our strategy, as explained in ref.\
\cite{PRD98}, is to locate the P-vortices
via center-projection on an uncooled lattice, count the number of
P-vortices piercing the minimal area of each loop, and then evaluate the
loops in the corresponding cooled, unprojected lattice.  Cooling tends
to thicken the core of center vortices \cite{PRD98}; according to our model 
this effect should simply extend Casimir scaling out to larger distances.

\begin{figure}[t!]
\centerline{\scalebox{.7}{\includegraphics{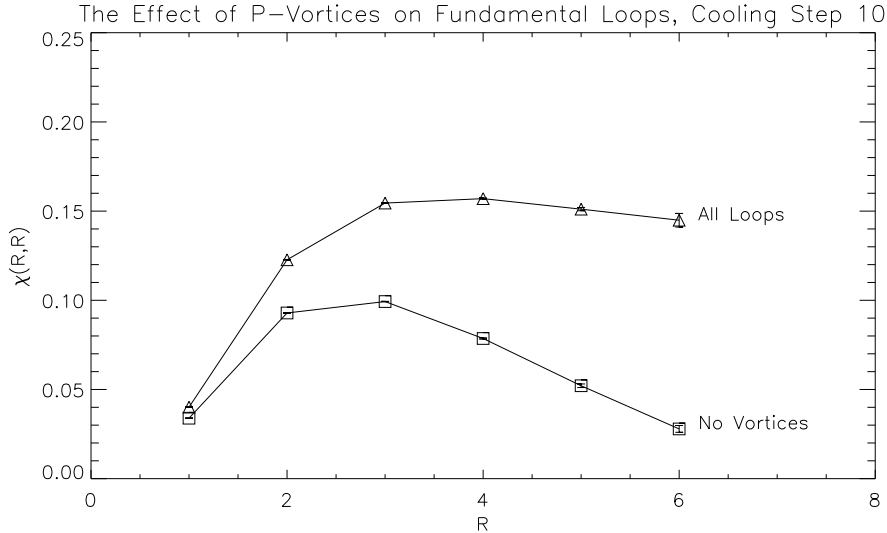}}}
\caption{Fundamental representation Creutz ratios on the cooled lattice,
at $\b=2.3$.  Results are shown for the full data set (triangles), and for 
the no-vortex loops (squares).}
\label{vtfund}
\end{figure}

\begin{figure}
\centerline{\scalebox{.7}{\includegraphics{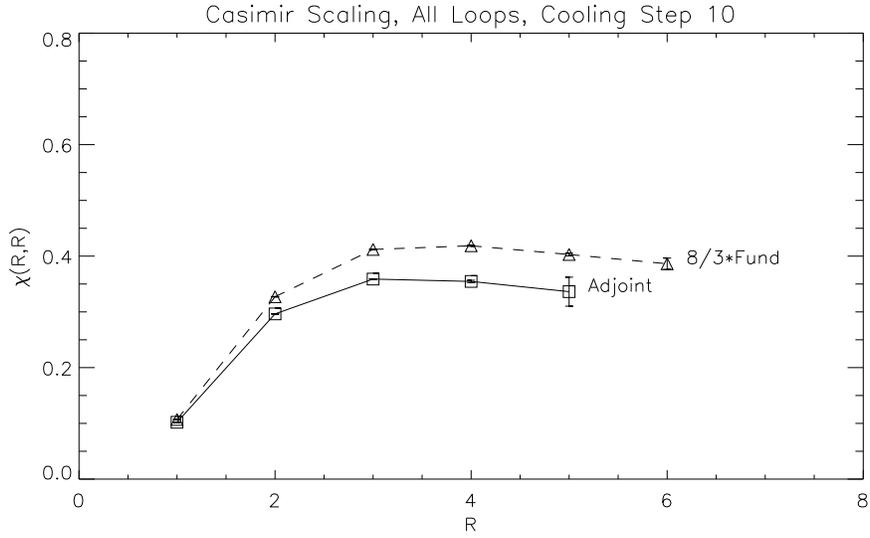}}}
\caption{All-loop adjoint Creutz ratios (squares) compared to 
the corresponding Casimir-rescaled (i.e. $\times 8/3$) data for
the all-loop fundamental Creutz ratios (triangles).}
\label{casall}
\end{figure}

\begin{figure}
\centerline{\scalebox{.7}{\includegraphics{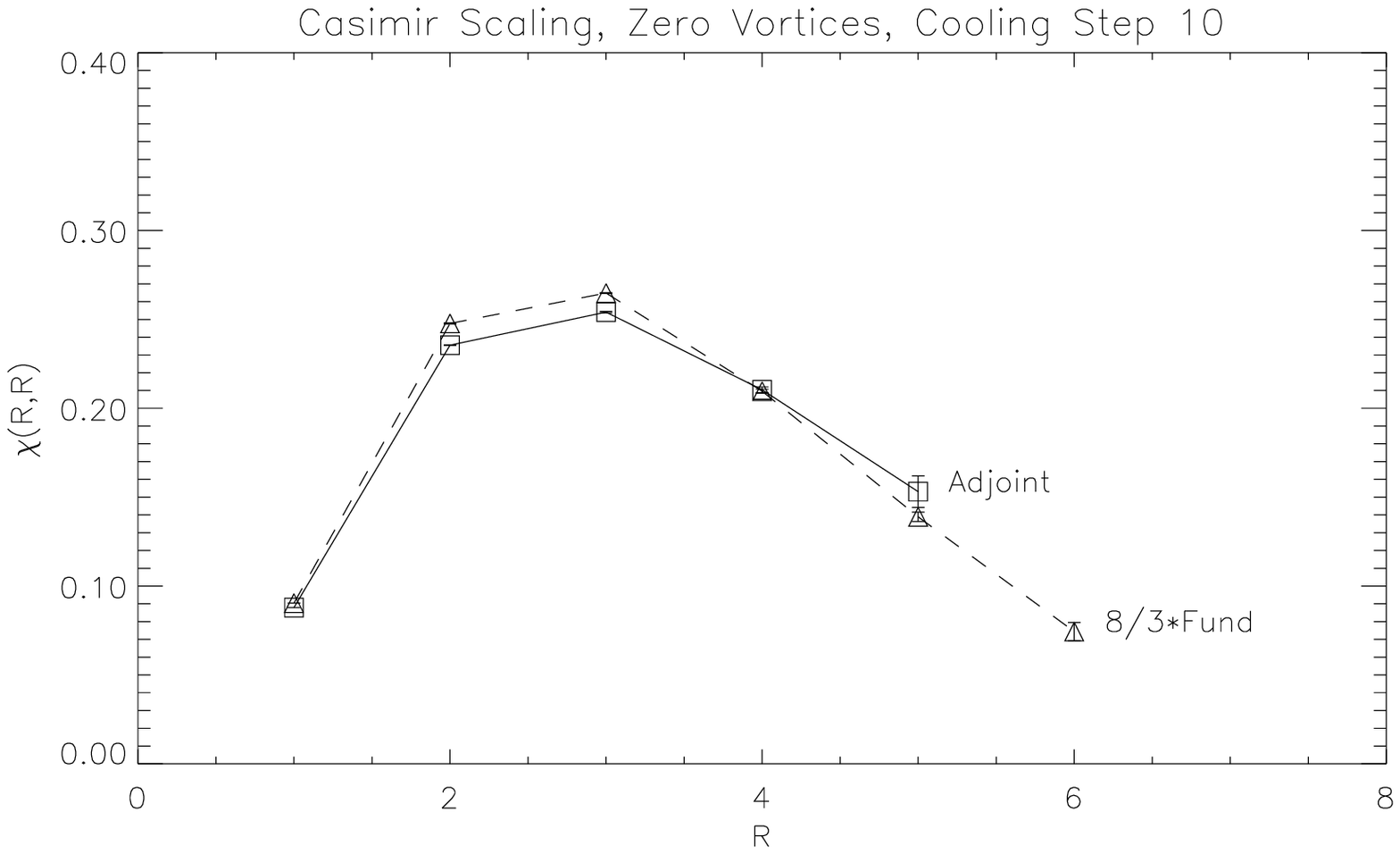}}}
\caption{Zero-vortex adjoint Creutz ratios (squares) compared to 
the corresponding Casimir-rescaled (i.e. $\times 8/3$) data for
the zero-vortex fundamental Creutz ratios (triangles).} 
\label{caszero}
\end{figure}
  
   Figure \ref{vtfund} shows our results for Creutz ratios in the
fundamental representation after 10 constrained cooling steps.
Data for this and the subsequent figures was taken from 760 configurations
separated by 100 update sweeps, on a $16^4$ lattice at $\b=2.3$, with loops 
evaluated after 10 constrained cooling sweeps of each configuration.  The 
upper line in Fig.\ \ref{vtfund} joins data points for 
the standard Creutz ratio $\chi_F(R,R)$, extracted from all fundamental Wilson 
loops of the appropriate sizes, while the lower line
joins data points for $\chi^0_F(R,R)$, extracted from only 
the zero-vortex fundamental Wilson loops $W_F^0(C)$.  Very similar data, 
up to $R=5$, was reported previously in ref.\ \cite{PRD98}.

   Figure \ref{vtfund} shows quite clearly, on a cooled lattice, what has 
also been found on uncooled lattices; namely, the no-vortex restriction
effectively eliminates the asymptotic string tension of Wilson loops 
\cite{PRD98,Zako}.  The full Creutz ratio varies only a little, in the
range from $R=3$ to $R=6$, and we expect it to converge to the 
nearby value of the asymptotic string tension (quoted as $\s=0.136(2)$ 
in ref.\ \cite{Bali}) as $R\ra \infty$. In sharp constrast, the zero-vortex 
ratio drops drastically in the same range, and is evidently tending 
to zero.    

   Figure \ref{casall} is an illustration of Casimir scaling on the
cooled lattice.  The solid line connects Creutz ratios $\chi_A(R,R)$ 
(squares) extracted from all loops in the adjoint-representation of 
appropriate size; for comparison we also display the corresponding data 
${8\over 3} \chi_F(R,R)$ (triangles) for fundamental Creutz ratios
rescaled by a factor of $8/3$.  If Casimir scaling were exact, these 
two sets of data points would coincide.  In fact, we find that 
$\chi_A(R,R)/\chi_F(R,R) \approx 2.25$ at $R=3,4,5$, rather than $2.66$.  
Because the signal is so much smaller for adjoint as compared to fundamental 
loops, we have only obtained meaningful data for the adjoint ratios up to 
$R=5$.  However, as far as we can tell from this limited data set, the 
all-loop adjoint Creutz ratios have stabilized around $R=3$ in parallel with 
the corresponding all-loop fundamental ratios.  The existence and 
(approximate) Casimir scaling of the adjoint string 
tension have also been seen in many previous studies \cite{Cas}.  

  The final (and crucial) figure is Fig.\ \ref{caszero}.  Here we show
the data for the zero-vortex adjoint Creutz ratios $\chi_A^0(R,R)$ (squares)
as compared to the zero-vortex fundamental Creutz ratios $\chi_F^0(R,R)$
(triangles), again rescaled by the factor of $8/3$.  There are two rather 
striking
features of this data: The first is that Casimir scaling of the zero-vortex
data is nearly exact.  The second is that the zero-vortex data for
the adjoint representation $\chi^0_A(R,R)$, like the zero-vortex data
for the fundamental representation, is falling rapidly towards zero.
It would be nice to have data at still larger $R$, particularly for the
adjoint zero-vortex loops, but in the data that we do
have there is not the slightest indication of the zero-vortex ratios 
stabilizing at a finite value, which would be required for the existence
of a non-zero string tension in the Casimir regime.
 
   The conclusion is that if we make a ``cut'' in the Monte-Carlo data
so that only zero-vortex loops, as defined above, are evaluated, 
then there is no sign of a finite string tension for either fundamental
\emph{or} adjoint loops in the Casimir-scaling regime. 
The simplest interpretation of this data is that 
center vortices give rise, not only to the fundamental, but also to the 
adjoint string tension in the Casimir regime.  This effect, in our opinion, 
is due to the mechanism proposed in ref.\ \cite{Us}.

\vspace{33pt}

\ni {\Large \bf Acknowledgements}

\bigskip

   We would like to thank J.\ Ambj{\o}rn for helpful discussions and comments,
and also for providing access to a fast workstation.

   This work was supported in part by Fonds zur F\"orderung der
Wissenschaftlichen Forschung P11387-PHY (M.F.), 
the U.S. Department of Energy under Grant No. DE-FG03-92ER40711
and Carlsbergfondet (J.G.), and the Slovak Grant Agency
for Science, Grant No. 2/4111/97 (\v{S}. O.).


\begin{thebibliography}{xx}
\bibitem{PRD98} L. Del Debbio, M. Faber, J. Giedt, J. Greensite, and 
{\v S}. Olejn\'{\i}k, hep-lat/9801027.
\bibitem{Zako} L. Del Debbio, M. Faber, J. Greensite, and 
{\v S}. Olejn\'{\i}k, in {\sl New Developments in Quantum Field Theory}, 
edited by P. Damgaard and J. Jurkiewicz (Plenum, New York, 1998), 
hep-lat/9708023; \\
L. Del Debbio, M. Faber, J. Greensite, and 
{\v S}. Olejn\'{\i}k, Phys. Rev. D55 (1997) 2298, hep-lat/9610005.
\bibitem{LR} K. Langfeld, O. Tennert, M. Engelhardt, and H. Reinhardt, 
hep-lat/9805002; hep-lat/9801030; \\
K. Langfeld, H. Reinhardt and O. Tennert, Phys. Lett. B419 (1998) 317,
hep-lat/9710068.
\bibitem{TK} T. Kov\'{a}cs and E. Tomboulis, Phys. Rev. D57 (1998) 4054,
hep-lat/9711009.
\bibitem{lat96}L. Del Debbio, M. Faber, J. Greensite, and 
{\v S}. Olejn\'{\i}k, Nucl. Phys. Proc. Suppl. 53 (1997) 141, 
hep-lat/9607053.
\bibitem{Cas1} L. Del Debbio, M. Faber, J. Greensite, and 
{\v S}. Olejn\'{\i}k, Phys. Rev. D53 (1996) 5891, hep-lat/9510028.
\bibitem{Us} M. Faber, J. Greensite, and {\v S}. Olejn\'{\i}k,
Phys.\ Rev.\ D57 (1998) 2603, hep-lat/9710039.
\bibitem{Cas} J. Ambj{\o}rn, P. Olesen, and C. Peterson, Nucl. Phys.
B240 [FS12] (1984) 198; 533; \\
C. Michael, Nucl. Phys. Proc. Suppl. 26 (1992) 417; Nucl. Phys. B259
(1985) 58; \\
N. A. Cambell, I. H. Jorysz, and C. Michael, Phys. Lett. B167 (1986) 91; \\
M. Faber and H. Markum, Nucl. Phys. Proc. Suppl. 4 (1988) 204; \\
M. M\"{u}ller, W. Beirl, M. Faber, and H. Markum, Nucl. Phys. Proc.
Suppl. 26 (1992) 423; \\
G. Poulis and H. Trottier, Phys. Lett. B400 (1997) 358, hep-lat/9504015.
\bibitem{Corn} J. M. Cornwall, Phys. Rev. D57 (1998) 7589, hep-th/9712248; \\
and in Proceedings of the {\sl Workshop on Non-Perturbative Quantum
Chromodynamics}, edited by K. A. Milton and M. A. Samuel (Birkhauser,
Boston, 1983).
\bibitem{cool} M. Campostrini, A. Di Giacomo, M. Maggiore, 
H. Panagopoulos, and E. Vicari, Phys. Lett. B225 (1989) 403.
\bibitem{Bali} G. S. Bali, C. Schlichter, and K. Schilling, Phys. Rev.
D51 (1995) 5165.

\end{thebibliography}
\end{document}